# Surfactant-Free Single Layer Graphene in Water

**Authors:** George Bepete,[1,2] Eric Anglaret,[3] Luca Ortolani,[4] Vittorio Morandi,[4] Alain Pénicaud[1,2*] and Carlos Drummond[1,2*]

**Affiliations:**

[1] CNRS, Centre de Recherche Paul Pascal (CRPP), UPR 8641, F-33600 Pessac, France.

[2] Univ. Bordeaux, CRPP, UPR 8641, F-33600 Pessac, France.

[3] Univ. Montpellier-II, Laboratoire Charles Coulomb (L2C), UMR CNRS 5521, F-34000 Montpellier, France.

[4] CNR IMM-Bologna, Via Gobetti 101, 40129 Bologna, Italy.

* drummond@crpp-bordeaux.cnrs.fr and penicaud@crpp-bordeaux.cnrs.fr

**Graphene and water do not mix. Liquid-phase exfoliation of graphite, motivated by the large number of potential applications of graphene [1] has been achieved by sonication or high-shear mixing, often introducing structural defects on the graphene lattice [2]. Best dispersions are a compromise between several factors such as number of layers (1 to 20 typically), lateral size (a few hundred nanometers) and concentration [3,4,5,6]. On the other hand, graphite intercalation compounds (GICs) can be readily exfoliated down to single layers (SLG) in aprotic solvents, yielding air- and moisture-sensitive graphenide (negatively charged graphene) solutions [7,8,9,10]. Here we show that homogeneous air-stable dispersions of SLG in water with no surfactant added can be obtained by mixing air-exposed graphenide solutions in tetrahydrofuran (THF) with *degassed* water and evaporating the organic solvent (Fig. 1). *In situ* Raman spectroscopy of this single layer**

**graphene in water (SLG$_{iw}$) shows all the expected characteristics of single layer, low-defect, graphene (Fig. 2). Accordingly, conductive films prepared from SLG in water exhibit a conductivity of up to 32 kS/m for a 15 nm thick film.**

In degassed water graphene re-aggregation is drastically slowed down due to the small inter-graphene attractive dispersive forces (a consequence of graphene two-dimensional character) and the stabilizing electrostatic repulsion. As has been reported before for many hydrophobic objects, (i.e. hydrocarbon droplets[11,12] or air bubbles[13]) graphene becomes electrically charged in water as a consequence of the spontaneous adsorption on its surface of OH$^-$ ions coming from graphenide oxidation and water dissociation. As two graphene flakes come together, they experience a repulsive force due to the overlap of their associated counterion clouds. Accordingly, graphene can be efficiently dispersed in water at a concentration of 0.16 g/L with a shelf life of a few months.

The pH values after graphene transfer to water is very revealing. While the system resulting from the mixture with non-degassed water (left vial of Fig. 1b) has a pH close to 11, stable graphene suspensions have a pH close to neutrality (pH between 7 and 8; right vial of Fig. 1b). As the same amount of OH$^-$ is produced in both cases after graphenide oxidation, the remarkable difference in pH is attributed to the adsorption of OH$^-$ on the suspended graphene flakes. This hypothesis is supported by the electrophoretic mobility and zeta potential $\zeta$ of the graphene flakes. Negative $\zeta$ values ($\zeta$ = -45 ± 5) were observed at neutral pH conditions; on the contrary, charge reversal was observed in acidic pH environment ($\zeta$ = +4 ± 2 at pH 4). It could be argued that this $\zeta$ variation is due to the reduction of pH below the pK$_a$ of functional groups dissociated at basic pH. To discard this hypothesis, we measured $\zeta$ of water-dispersed graphene in presence of tetraphenylarsonium chloride, Ph$_4$AsCl which contains a hydrophobic cation known to readily adsorbs on hydrophobic surfaces [14]. As reported in Table 1, we observed a progressive increase in $\zeta$ with increasing concentration of the hydrophobic cation, with charge reversal at sufficiently large cation concentrations.

| [Ph$_4$AsCl] (mM) | $\zeta$ (mV) |
|---|---|
| 0 | -45 ± 5 |
| 1 | -21 ± 4 |
| 2 | -10 ± 4 |
| 5 | +5 ± 2 |

**Table 1.** Zeta potential of graphene flakes dispersed in water for different concentrations of Ph$_4$AsCl.

Several hypotheses have been advanced to explain the ionic adsorption on hydrophobic surfaces often observed; favorable entropy changes due to partial release of ionic hydration layer upon adsorption [15], asymmetry of water ions [16], dispersion interactions related to ionic polarizability and ionic-induced decrement of water polarization fluctuations [17] are some examples discussed in the literature. For the particular case of graphene, the adsorption is also likely to be promoted by its conducting character.

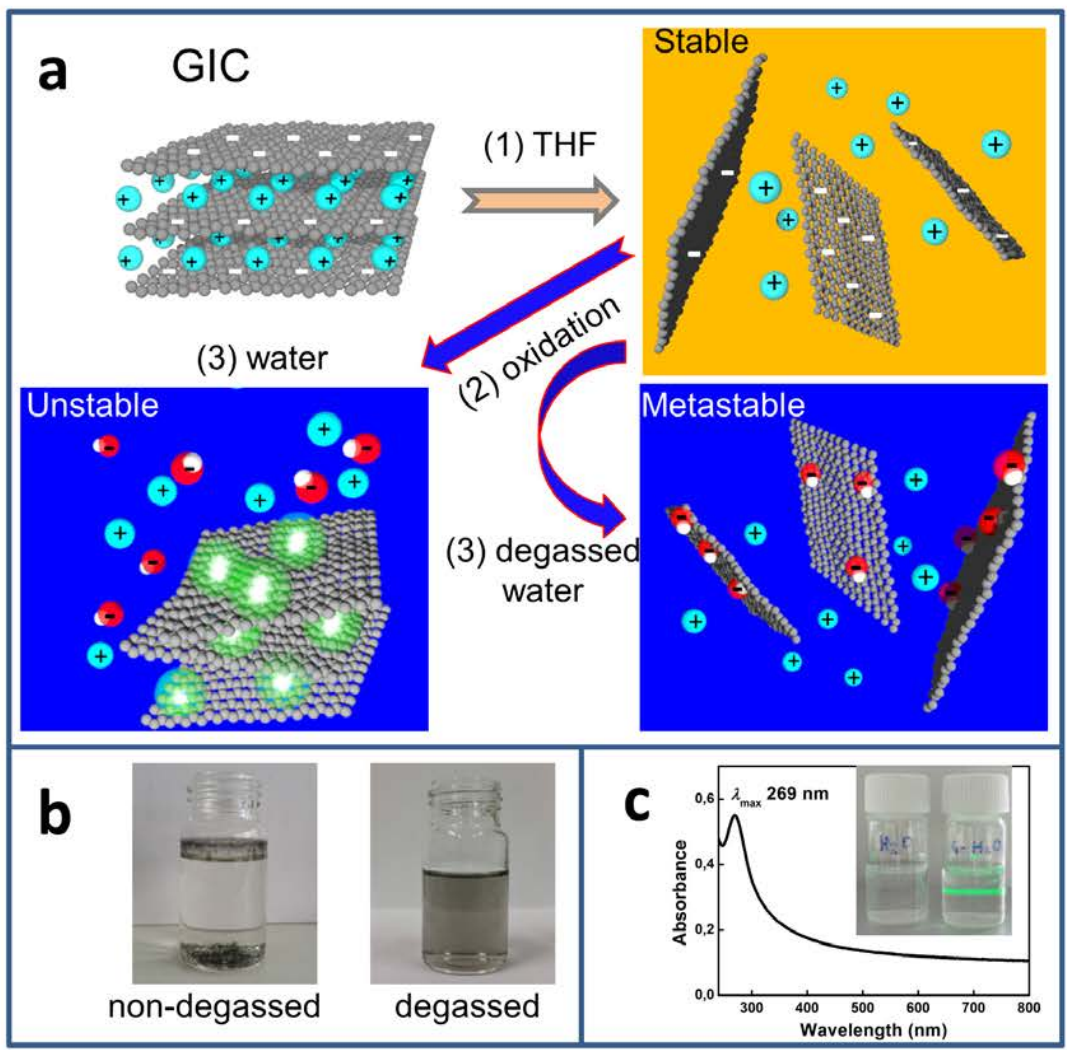

**Figure 1.** (a) Preparation of $SLG_{iw}$. $KC_8$ is solubilized in THF under inert atmosphere as single layer graphenide polyions. Graphenide ions are then oxidized back to graphene in THF by air exposure and immediately transferred to degassed water. Upon air exposure, graphenide reduces oxygen to superoxide anion[18] (that eventually yields hydroxide anion), while graphenide turns to neutral graphene[10], with some minor functionalization (vide infra). Stability of $SLG_{iw}$ is determined by the interaction between the individual graphene plates. In regular laboratory conditions, gases dissolved in water (about 1 mM) adsorb on the graphene surface, inducing long-range attractive interaction between the dispersed objects and promoting aggregation (a, bottom left, gas bubbles and ions are not at scale). On the contrary, if water is degassed (removing dissolved gases) water-ions readily adsorb on the graphene surface, conferring a certain charge to the dispersed objects. The repulsive electrostatic interaction favors the stability of the dispersed material (b) Left vial: mixture of graphene in THF after addition to water which was not degassed. The aqueous dispersion is not stable and black aggregates visible to the eye begin to form a few minutes after mixing. Right vial: stable dispersion of graphene in degassed water after THF evaporation. No evidence of aggregation is observed after several months of storage at room temperature (c) UV-visible absorption spectrum shows an absorption peak at 269 nm (4.61 eV), the exact wavelength reported for the absorption of a single layer of graphene on a substrate [19]. Inset: laser goes through water unscattered (left) whereas a graphene dispersion (right) shows Tyndall effect due to light scattering by large graphene flakes.

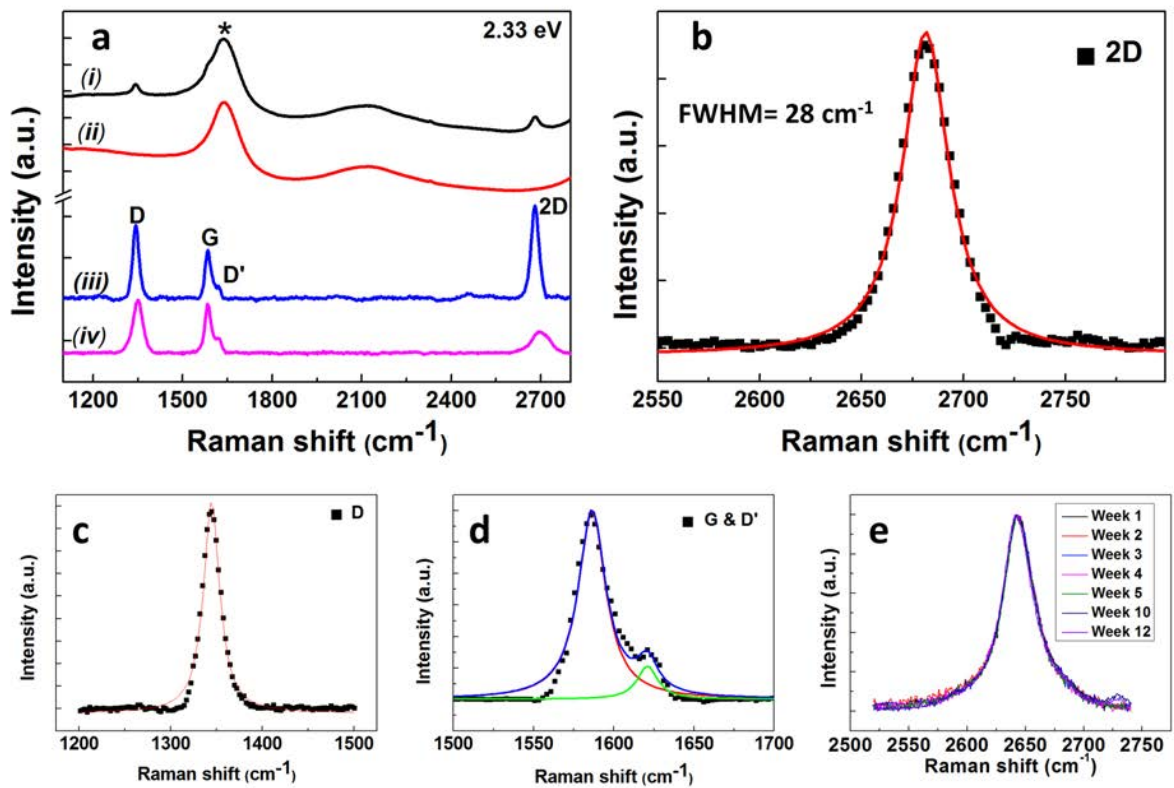

**Figure 2.** Graphene spectra have been obtained by subtraction of the spectrum of pure water from that of the graphene dispersions measured in the same cuvette, normalized on the bending peak of water (starred in (a)). (a) From top to bottom : Raman spectra of (i) $SLG_{iw}$ dispersion, (ii) water, (iii) graphene after subtraction of water, at 2.33 eV. For comparison, a spectrum of a sonicated sodium cholate few layer graphene dispersion is presented in (iv) (prepared according to experimental details of ref [6]. (b – d) Typical fits of the 2D, D, G and D' peaks of SLG in water at 2.33 eV. The slight asymmetry in the fit of the 2D line is due to imperfections in the water background subtraction. (e) Raman 2D band as a function of time (at 1.94 eV) showing excellent time stability; the corresponding full spectra are presented in supplementary Figure S1).

Raman spectroscopy has been used as a powerful tool to study graphene samples, to determine number of layers, stacking sequence in the case of multiple layers, doping, amount and nature of defects [20]. The Raman spectrum of $SLG_{iw}$, (Figure 2 & Table 2) shows typical features of SLG such as a narrow, symmetrical, intense 2D (also called G') band of full width at half maximum (FWHM) below 30 cm$^{-1}$. Good fits of the 2D, D, G and D' peaks are obtained using single Lorentzian lines (Figure 2 b-d). It is interesting to compare the Raman spectrum of $SLG_{iw}$ with other aqueous dispersions, such as sonication aided sodium cholate (SC) suspensions prepared according to ref [6] (spectrum (iv) in Fig. 2a). Quality of the exfoliation is readily apparent from the much sharper and more intense 2D band for SLG$_{iw}$ (spectrum (iii)) while the D band is only slightly enhanced compared to sonication-aided dispersions (spectrum iv). Finally, stability of these aqueous dispersions is addressed in Fig. 2e where the temporal evolution of the Raman 2D band is presented. No apparent change can be seen after few months of storage. Likewise, light scattering experiment show no change over a few month period (SI, Figure S2). As air re-dissolution in water is known to happen on a short time scale (hours at most), stability of SLG$_{iw}$ with time shows that once adsorbed, the OH- ions are not displaced by dissolved gas.

| Excitation Energy (eV) | D | | G | | D' | | 2D | | $I_D/I_G$ | $I_D/I_{D'}$ | $I_{2D}/I_G$ |
|---|---|---|---|---|---|---|---|---|---|---|---|
| | Pos | FWHM | Pos | FWHM | Pos | FWHM | Pos | FWHM | | | |
| 2.33 | 1345 | 27 | 1586 | 21 | 1620 | 16 | 2681 | 28 | 1.5 | 9.0 | 2.0 |

**Table 2.** Raman characterization: Peaks position (cm$^{-1}$), full width at half maximum (FWHM, cm$^{-1}$) and relevant intensity ratios at excitation energy 2.33 eV. Similar results for different excitation energies are presented in Supplementary Table S1.

**Single-layeredness**: A key Raman signature of single layer graphene (SLG) is the intensity, shape and width of the 2D (G') band. Multilayer, AB stacked (Bernal) few layer graphene shows a 2D band with a complex shape fitted by a number of Lorentzian lines [21]. Turbostratic graphite, i.e. graphite with uncorrelated graphene layers, shows a single Lorentzian 2D band with a FWHM of 50 cm$^{-1}$ [21]. On the contrary, the intense 2D band of supported SLG can be well fitted by using single Lorentzians of FWHM between 20 and 35 cm$^{-1}$ [22], and suspended graphene shows a 2D FWHM of 24 +/- 2 cm$^{-1}$ [23]. Therefore, the observed 2D band at 2681 cm$^{-1}$ (at 2.33 eV) with an intensity twice that of the G band, a pure Lorentzian shape, a FWHM of 28 cm$^{-1}$ (and a dispersion of 119 cm$^{-1}$/eV) strongly supports that SLG$_{iw}$ contains mainly, if not only, single layer graphene. The other characteristics of the Raman spectra (Table 2) are all in agreement with the literature for SLG.

Deposits were also made from SLG$_{iw}$ (Figure 3). Fig. 3a, 3b and Supplementary Figure S3 show AFM topographic images. Natural graphite contains domains of different sizes; small and large flakes will be present. If the flakes are too large (typically larger than few μm) they will likely fold on themselves (as evidenced by TEM results) and will be difficult to image by AFM. For AFM, a region was chosen with with many small flakes (Figure 3a) to be able to build a meaningful thickness distribution (inset of Fig. 3a) showing that single layer graphene is being produced. Other AFM micrographs (Fig. 3b and supporting information) were chosen to show the different sizes of mostly single layer graphene that are obtained. Statistics on ca 150 flakes (Inset to Fig. 3a) show that all objects have a thickness consistent with single (0.34 nm) or double (0.68 nm) layer, with a majority of single layers. AFM results are corroborated by TEM. Fig. 3c (and Supplementary Figures S4-S5) reveals the crumpled geometry of the flakes after deposition. Electron diffraction analysis (Supplementary Figure S4) confirms the graphitic structure of the deposited material, while the degree of exfoliation of the flakes can be estimated by carefully analyzing folded edges. Unfortunately, the crumpled and multiply folded nature of the deposited material prevents a precise determination of the thickness of each flake. Nevertheless the uniformity of the TEM image contrast reveals homogeneous exfoliation, and the abundance of folds showing only one (002) graphite fringe in the HRTEM image, definitely confirms Raman and AFM findings of extensive monolayers in the produced material.

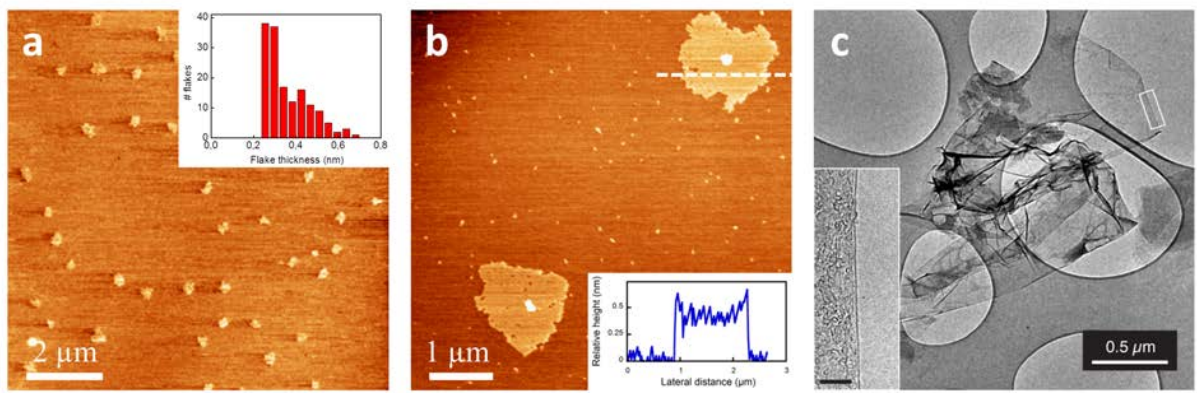

**Figure 3.** Characterization of deposits from $SLG_{iw}$. Deposits were made by dip coating. (a) & (b) Topographic images on mica by AFM show homogeneous thickness of the deposited graphene flakes. Inset to a: thickness distribution. Inset to b: height profile along the dashed line. (c) TEM micrograph of a flake deposited from the liquid solution over the TEM grid. (inset) High-resolution TEM (HRTEM) of a folded flake. The number of graphite (002) fringes visible at the edge allows a direct measurement of the local number of graphene layers (monolayer fold). Scale bar corresponds to 5 nm. Supplementary Figure S5 shows additional results of the TEM characterization of the flake borders.

Two requirements are necessary to formulate SLG-liquid dispersions: the production of SLG and its transfer to the liquid matrix. The practical value of the obtained dispersion will be ultimately governed by its stability. Three factors converge to promote the stability of $SLG_{iw}$, as can be ascertained from the graphene-graphene energy of interaction (Figure 4): the adsorption of $OH^-$ ions on graphene, the reduction of hydrophobic interaction in the absence of dissolved gases, and the relatively weak van der Waals interactions between SLG by virtue of their two dimensional character. The stability of $SLG_{iw}$ is governed by the difference between the repulsive electrostatic interaction and the destabilizing attractive forces (dispersion and hydrophobic). Graphene flakes experience attractive hydrophobic interaction in water as a consequence of their disruptive effect on the water hydrogen bond network [24,25]. It has also been argued that, in presence of dissolved gases, long-range capillary attraction appears, due to nanobubbles adsorbed on hydrophobic surfaces or to a zone of depleted density close to the interfaces. When gases are thoroughly removed, the range of this interaction is substantially reduced, as has been observed by direct measurement of surface forces in a number of studies[26,27]. Attractive dispersion interaction is another destabilizing contribution to the inter-flakes interaction. The van der Waals interaction energy (per unit area) $W_{vdW}$, between flakes of thickness $a$ at a separation $D$, can be estimated as $W_{vdW} = -\frac{A_{Ham}}{12\pi}\left[\frac{1}{D^2} - \frac{2}{(D+a)^2} + \frac{1}{(D+2a)^2}\right]$, where $A_{Ham}$ is the Hamaker coefficient for the particular combination of materials (graphene-graphene in water). For thick objects $W_{vdW} \sim 1/D^2$ and the value of $a$ is inconsequential. On the contrary, the effect of the finite thickness is notorious when $D$ is comparable or larger than $a$ [28]. There are two important consequences of this attractive force. First, few-layer objects will be less stable than SLG: the increasing dispersion interaction substantially reduces the energy barrier to flake aggregation when the thickness of the dispersed flakes increases (Fig.4b). More interestingly, the secondary attractive potential energy minimum —normally observed as a consequence of the prevalence of dispersive over electrostatic interaction at large separations— is not present for SLG, due to the fast decay of the attractive interaction (Fig.4c). Hence, loose flocculation, a factor responsible for instability of many micron-size object dispersions, is absent for the case of charged SLG in water. A more detailed discussion about graphene inter-flake interaction is presented as supplementary information.

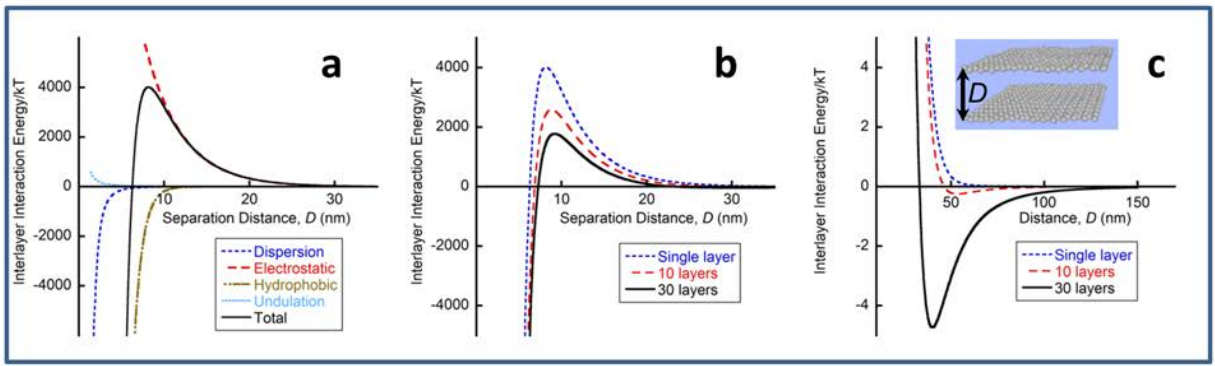

**Figure 4.** (a) The interplate interaction energy $W$ can be estimated by adding up the different contributions, as discussed in the SI. A non-monotonic $W$ vs. $D$ behavior, with an energy barrier slowing down the aggregation, is obtained from the competition between attractive and repulsive interactions; the larger the energy barrier the more stable the graphene dispersion will be. (b) The attractive component quickly increases with the number of layers of the dispersed objects —shifting from 2D to 3D objects— reducing the energy barrier that assures the dispersion stability. (c) The secondary minimum, observed at large separations for few layer flakes, is responsible for the flocculation and poor dispersibility of thin graphite. This minimum is not observed for SLG. Flake lateral size 0.5 µm T 298 K.

Reports of "graphene" dispersions abound. They actually show a distribution of thickness, ranging from 1 to 20 layers in the best cases [3,6,29]. By dispersing graphite with the help of mechanical energy, one goes against thermodynamics, to break apart the efficient packing of graphene in graphite. Hence, the resulting dispersion *has to* be a statistical distribution of thicknesses with single layer flakes forming the tail of that distribution. Since we start from fully exfoliated graphenide flakes, all that is needed is an energy barrier to circumvent graphene re-aggregation. Degassed water affords that barrier without the need for any additive, apart from the $OH^-$ ions. Although a large number of reports claim exfoliation of graphite into graphene, Raman characterization of those dispersions *in situ* is rare. One of the very few Raman spectra in liquid of a graphene dispersion shows a symmetrical, Lorentzian shaped, 2D band with a FWHM of 44 $cm^{-1}$, attributed to turbostratically packed few layer graphene [30]. $SLG_{iw}$, on the contrary, shows a clear Raman signal of SLG in a liquid. At first sight, D band appears large. However, one is not measuring a single flake but a large number of them, of all sizes and orientations. Edges will naturally have a large contribution although they do not fully account for the intensity of the D band, as some $sp^3$ defects have been created in the process. However, as quantified according to [31], the defect concentration in $SLG_{iw}$ amounts to 300-600 ppm only (see Table S1 and the details of the calculation in Supp. Inf.). Further proof of the low amount of defects is given by X-ray photoelectron spectroscopy (XPS) analysis of the films showing minor widening on the high energy side of the C1s peak (see sup. Info). Actually, the minute (and controllable [32]) amount of defects in $SLG_{iw}$ represents an opportunity for further functionalization e.g. with responsive or biologically relevant functions. Finally, the exceptional exfoliation level of $SLG_{iw}$ and low defect level is reflected in the conducting properties of materials made from it: Conductive coatings prepared by filtering SLG$_{iw}$ show average conductivities of 7 and 20 kS/m after annealing at 200 and 500 °C respectively, for films of only 15 or 30 nm thickness (see supp. Info). The best device exhibited a sheet resistance of 2100 Ohm/sq (at 60 % transparency), a value to be compared to the best of their kind within RGO films, exhibiting sheet resistances of 840 Ohm/sq and 19.1 kOhm/sq for flakes of respective mean size of 7000 and 200 μm$^2$ flakes[33]. Average flake area in our films is 1 μm$^2$ that should translate into quite resistive films if it were not for the quality of the flakes. The equivalent bulk conductivity of this film is 32 kS/m opening exciting perspectives for conductive coatings and composite applications of graphene films.

Implications of this work are four-fold: (i) graphene can be efficiently dispersed in water, as true single layers, with no additives, at a concentration of 0.16 g/L and with shelf life of several months. This remarkable feat being due to graphene 2D character, SLG$_{iw}$ might well find use to produce additive free aqueous dispersions of other 2D materials (ii) As has been the case for graphene obtained by mechanically exfoliation of graphite, the intensity, shape and width of the Raman 2D band are proposed as very sensitive quality parameters of graphene aqueous dispersions and composites. (iii) By providing true SLG in water, a vast amount of potential applications can be readily envisioned such as drug carriers, toxicology studies, biocompatible devices, composites, patterned deposits exploiting the superior electrocatalytic performance of carbon surfaces in general and of graphene in particular, impregnation of 3D architectures for supercapacitors and other energy related applications. (iv) SLG$_{iw}$ brings new experimental evidence regarding the hydrophobic surface / water interaction.

**Methods**

**1. Preparation of graphenide solution.** Under inert atmosphere, 108 mg of $KC_8$ were dispersed in 18 mL of distilled THF and this mixture was tightly sealed and mixed for 6 days with a magnetic stirrer (900 rpm). After stirring, the solution was left to stand overnight to allow non-dissolved graphitic aggregates to form and settle at the bottom. The mixtures were centrifuged in 10 mL glass vials at 3000 rpm for 20 minutes. The top two thirds of the solution were extracted with a pipette and retained for use.

**2. Transfer of graphene from THF to water.** Under ambient atmosphere, the centrifuged graphenide THF solution was left exposed to air for 1 minute and then added carefully to previously degassed water and left open to let THF evaporate for two days. Degassing was achieved by subjecting the water to mild agitation (using a carefully cleaned magnetic Teflon bar stirrer to induce the nucleation of gas bubbles) under pressure of 0.2 mbar for 30 min. Then the air pressure was gently increased back to atmospheric pressure. 7 mL of degassed water were transferred to a 20 mL glass vial; graphenide solution was air exposed for 1 minute and was added drop-wise to the degassed water with gentle stirring using a stainless steel needle. The vial was left open in a dust-free environment to allow THF evaporation at room temperature whilst stirring gently with a steel needle every hour for the first ten hours and occasionally thereafter to yield a slightly dark dispersion of graphene in water. Different graphene concentrations in water were obtained by varying the ratio of THF graphenide solution and water. The dispersions were characterized using absorption spectroscopy, Raman spectroscopy, dynamic light scattering. The yield of dispersed SLG vs. starting graphite is 4 %.

**3. Electrophoretic mobility** of graphene in $SLG_{iw}$ was measured using a Zetacompact Z8000 (CAD Instrumentation, France). An electric field of 8.95 V/cm was applied and graphene mobility was measured by direct particle tracking. Due to the large concentration of graphene in $SLG_{iw}$, the suspensions were diluted 100 times before the measurements. Zeta potential of graphene flakes $\zeta$ was calculated from its electrophoretic mobility applying the Smoluchowski equation [34].

**4. Raman spectroscopy** was performed on an Xplora spectrometer from Horiba-Jobin-Yvon at 2.33 eV excitation energy (532 nm laser wavelength) using a macro sample holder containing a cuvette filled up with $SLG_{iw}$ (1 cm pathway). Peak positions were calibrated using the $T_{2g}$ peak of silicon (520.5 cm$^{-1}$) and the G band of HOPG (1582 cm$^{-1}$).

**5. Dynamic light scattering:** The size and state of aggregation of the SLG in $SLG_{iw}$ was determined by Dynamic Light Scattering, DLS (ALV 5000 CGS). The autocorrelation function of the scattering intensity, $g^2(q;\tau)$, is exquisitely sensitive to the size of particles in the dispersion. No significant changes were observed in $g^2(q;\tau)$ after several weeks of storage of $SLG_{iw}$ at room temperature, as can be observed in Supplementary Figure S2. Mean lateral size obtained is 0.9 micrometer.

**6. AFM deposits:** Deposits were obtained by dip coating a freshly cleaved mica substrate in $SLG_{iw}$ by itself or containing 1 mM $AsPh_4Cl$ salt. The positively charged $AsPh_4Cl$ salt ions adsorb on the graphene flakes, conferring them a positive charge (as verified by zeta potential measurements) and improving adsorption. The deposits were rinsed with distilled water followed by blow drying with dry $N_2$ gas. Topography micrographs were measured using an AFM Icon (Bruker).

**7. Transmission Electron Microscopy**: SLG$_{iw}$ was drop-cast on holey carbon grids for TEM characterization. Structural and morphological characterization of the material has been performed on FEI Tecnai F20 ST transmission electron microscope (TEM), operated at 120 kV of accelerating voltage to reduce the beam damage on the graphene, while preserving the resolution to image (0,0,2) graphite fringes for the measurement of the local thickness on folded edges. Local elemental analysis has been performed in-situ in the TEM using an energy dispersion X-ray spectrometer (EDX).

**References and Notes:**

1.  Bonaccorso, F. *et al.* Graphene, related two-dimensional crystals, and hybrid systems for energy conversion and storage. *Science (80-. )*. **347,** 1246501–1246501 (2015).

2.  Cravotto, G. & Cintas, P. Sonication-Assisted Fabrication and Post-Synthetic Modifications of Graphene-Like Materials. *Chem. - A Eur. J.* **16,** 5246–5259 (2010).

3.  Paton, K. R. *et al.* Scalable production of large quantities of defect-free few-layer graphene by shear exfoliation in liquids. *Nat. Mater.* **13,** 624–30 (2014).

4.  He, P. *et al.* Processable Aqueous Dispersions of Graphene Stabilized by Graphene Quantum Dots. *Chem. Mater.* **27,** 218–226 (2015).

5.  Ciesielski, A. & Samorì, P. Graphene via sonication assisted liquid-phase exfoliation. *Chem. Soc. Rev.* **43,** 381–98 (2014).

6.  Lotya, M., King, P. J., Khan, U., De, S. & Coleman, J. N. High-concentration, surfactant-stabilized graphene dispersions. *ACS Nano* **4,** 3155–62 (2010).

7.  Pénicaud, A. & Drummond, C. Deconstructing graphite: graphenide solutions. *Acc. Chem.*

*Res.* **46,** 129–37 (2013).

8. Milner, E. M. *et al.* Structure and morphology of charged graphene platelets in solution by small-angle neutron scattering. *J. Am. Chem. Soc.* **134,** 8302–8305 (2012).

9. Catheline, A. *et al.* Solutions of fully exfoliated individual graphene flakes in low boiling point solvents. *Soft Matter* **8,** 7882 (2012).

10. Englert, J. M. *et al.* Functionalization of graphene by electrophilic alkylation of reduced graphite. *Chem. Commun.* **48,** 5025 (2012).

11. Pashley, R. M. Effect of Degassing on the Formation and Stability of Surfactant-Free Emulsions and Fine Teflon Dispersions. *J. Phys. Chem. B* **107,** 1714–1720 (2003).

12. Carruthers, J. C. The electrophoresis of certain hydrocarbons and their simple derivatives as a function of p H. *Trans. Faraday Soc.* **34,** 300 (1938).

13. Zimmermann, R., Freudenberg, U., Schweiß, R., Küttner, D. & Werner, C. Hydroxide and hydronium ion adsorption — A survey. *Curr. Opin. Colloid Interface Sci.* **15,** 196–202 (2010).

14. Siretanu, I., Chapel, J., Bastos-González, D. & Drummond, C. Ions-Induced Nanostructuration: Effect of Specific Ionic Adsorption on Hydrophobic Polymer Surfaces. *J. Phys. Chem. B* (2013). doi:10.1021/jp400531x

15. Noah-Vanhoucke, J. & Geissler, P. L. On the fluctuations that drive small ions toward, and away from, interfaces between polar liquids and their vapors. *Proc. Natl. Acad. Sci. U. S. A.* **106,** 15125–30 (2009).

16. Kudin, K. N. & Car, R. Why are water-hydrophobic interfaces charged? *J. Am. Chem. Soc.* **130,** 3915–9 (2008).

17. Gray-Weale, A. & Beattie, J. K. An explanation for the charge on water's surface. *Phys. Chem. Chem. Phys.* **11,** 10994–11005 (2009).

18. Stinchcombe, J., Penicaud, A., Bhyrappa, P., Boyd, P. D. W. & Reed, C. a. Buckminsterfulleride(1-) salts: synthesis, EPR, and the Jahn-Teller distortion of C60-. *J. Am. Chem. Soc.* **115,** 5212–5217 (1993).

19. Kravets, V. G. *et al.* Spectroscopic ellipsometry of graphene and an exciton-shifted van Hove peak in absorption. *Phys. Rev. B* **81,** 155413 (2010).

20. Ferrari, A. C. & Basko, D. M. Raman spectroscopy as a versatile tool for studying the properties of graphene. *Nat. Nanotechnol.* **8,** 235–46 (2013).

21. Malard, L. M., Pimenta, M. a., Dresselhaus, G. & Dresselhaus, M. S. Raman spectroscopy in graphene. *Phys. Rep.* **473,** 51–87 (2009).

22. Wang, Y. ~Y. & Et Al. Raman Studies of Monolayer Graphene: The Substrate Effect. *J. Phys. Chem. C* **112,** 10637–10640 (2008).

23. Berciaud, S., Ryu, S., Brus, L. E. & Heinz, T. F. Probing the intrinsic properties of exfoliated graphene: Raman spectroscopy of free-standing monolayers. *Nano Lett.* **9,** 346–52 (2009).

24. Israelachvili, J. N. in *Intermol. Surf. Forces* 205–222 (Elsevier, 2011). doi:10.1016/B978-0-12-375182-9.10011-9

25. Chandler, D. Interfaces and the driving force of hydrophobic assembly. *Nature* **437,** 640–7 (2005).

26. Considine, R. F., Hayes, R. a. & Horn, R. G. Forces Measured between Latex Spheres in Aqueous Electrolyte: Non-DLVO Behavior and Sensitivity to Dissolved Gas. *Langmuir* **15,** 1657–1659 (1999).

27. Meyer, E. E., Rosenberg, K. J. & Israelachvili, J. Recent progress in understanding hydrophobic interactions. *Proc. Natl. Acad. Sci. U. S. A.* **103,** 15739–15746 (2006).

28. Parsegian, V. A. *Van der Waals Forces A Handbook for Biologists, Chemists, Engineers, and Physicists*. (Cambridge University Press, 2005).

29. Kovtyukhova, N. I. *et al.* Non-oxidative intercalation and exfoliation of graphite by Brønsted acids. *Nat. Chem.* **6,** 957–963 (2014).

30. Mary, R., Brown, G., Beecher, S. & Torrisi, F. 1.5 GHz picosecond pulse generation from a monolithic waveguide laser with a graphene-film saturable output coupler. *Opt. Express* **21,** 11 (2013).

31. Cançado, L. G. *et al.* Quantifying defects in graphene via Raman spectroscopy at different excitation energies. *Nano Lett.* **11,** 3190–6 (2011).

32. Schäfer, R. a *et al.* On the way to graphane-pronounced fluorescence of polyhydrogenated graphene. *Angew. Chem. Int. Ed. Engl.* **52,** 754–7 (2013).

33. Zhao, J., Pei, S., Ren, W., Gao, L. & Cheng, H. M. Efficient preparation of large-area graphene oxide sheets for transparent conductive films. *ACS Nano* **4,** 5245–5252 (2010).

34. Robert J. Hunter. *Foundations of Colloid Science*. (Oxford University Press, 2001).

## Acknowledgements

Support from the Agence Nationale de la Recherche (GRAAL) and Linde Corp. is acknowledged. AP thanks Nacional de Grafite (Brazil) for a gift of natural graphite. This work has been done within the framework of the GDR-I 3217 ''graphene and nanotubes''.

## Author Contributions

G.B. prepared and characterized $SLG_{iw}$, G.B, E.A, A.P & C.D. analyzed the experimental data and wrote the manuscript. L.O & V. M made the TEM analysis.

## Additional information

Correspondence and requests for materials should be addressed to C.D. (drummond@crpp-bordeaux.cnrs.fr) or A.P. (penicaud@crpp-bordeaux.cnrs.fr).

**Competing financial interests**

The authors declare no competing financial interests.

**Supplementary Information**

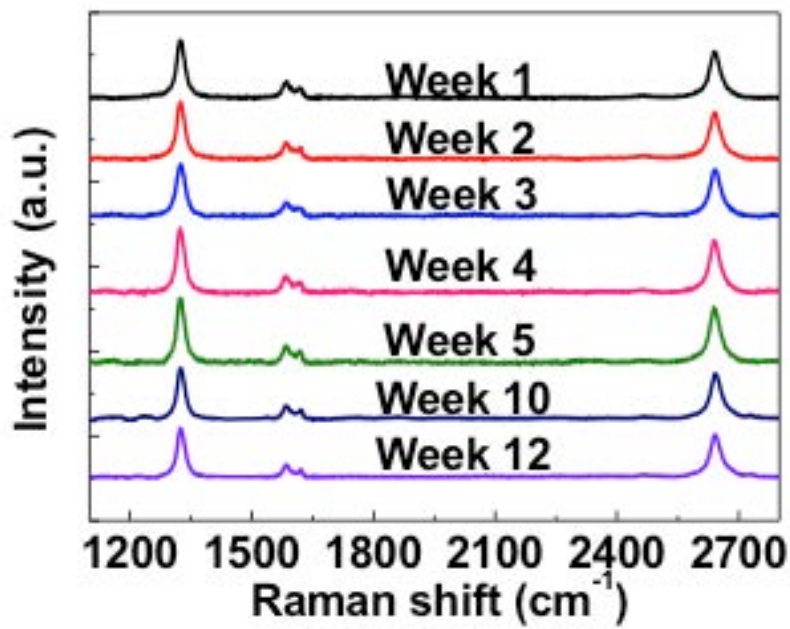

**Figure S1.** Full Raman spectra (at 1.94 eV) of $SLG_{iw}$ at different times after preparation. No evolution of the spectrum of $SLG_{iw}$ was observed over 12 weeks of storage.

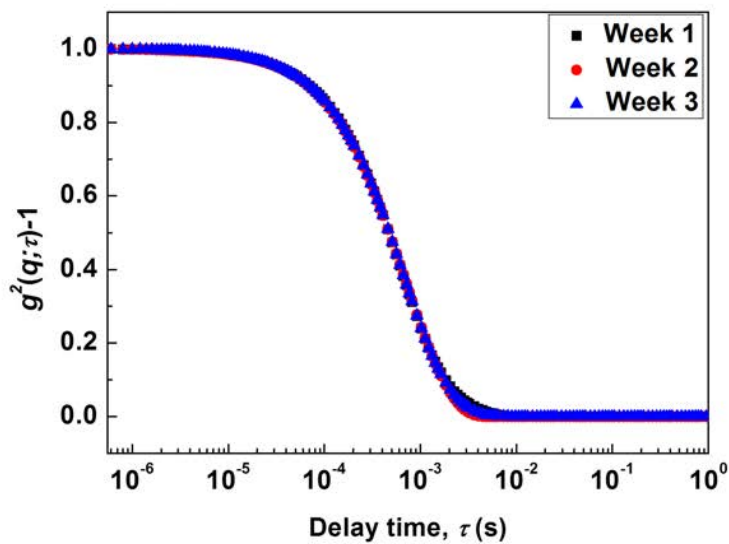

**Figure S2.** Scattering intensity autocorrelation function ($g^2(q;\tau)$) of SLG aqueous dispersion measured at different times of storage. Scattering angle $\theta = 90°$.

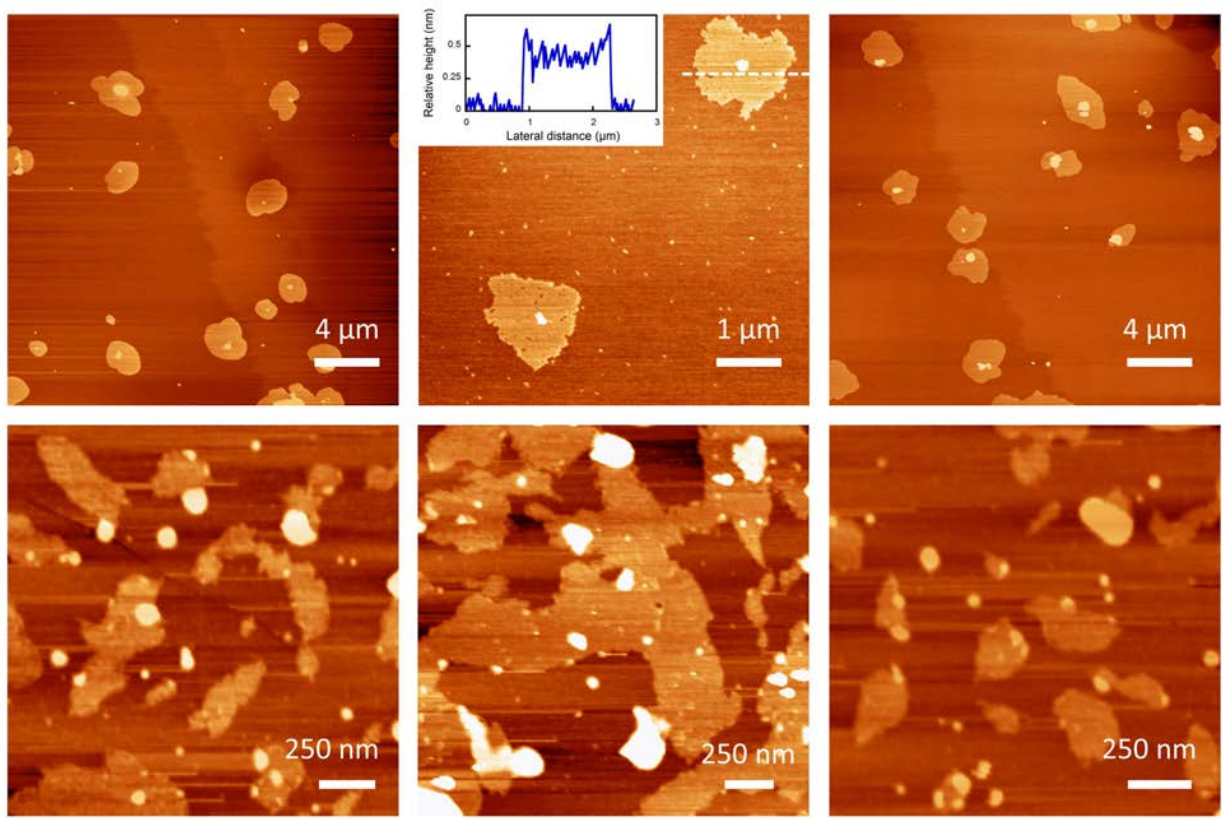

**Figure S3.** Additional AFM images. The first raw correspond to deposits from graphene-tetraphenyl arsonium $SLG_{iw}$; second raw corresponds to water-only $SLG_{iw}$. All heights are consistent with single layer graphene.

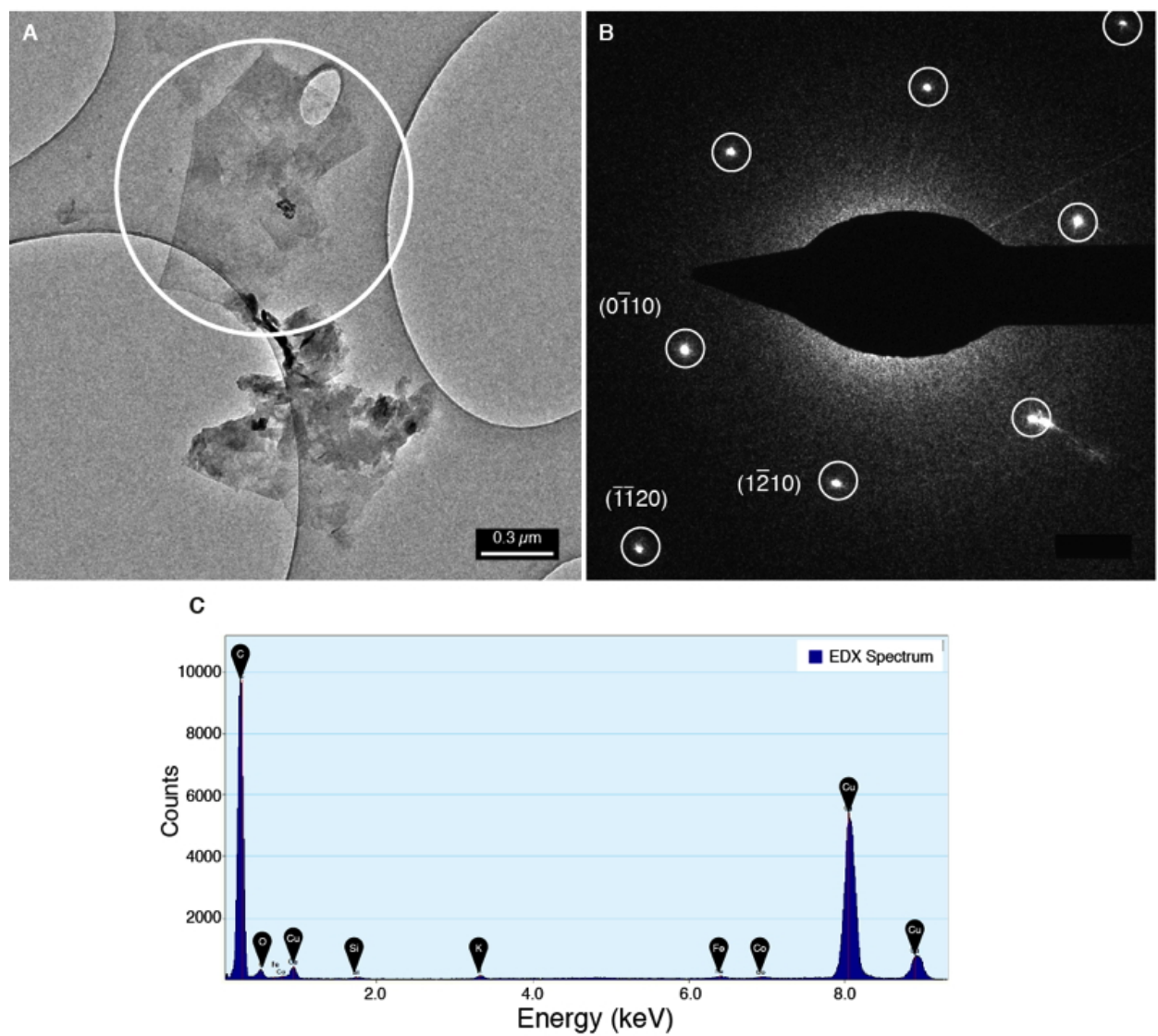

**Figure S4:** Additional TEM characterization. A) TEM micrograph of a graphene flake over the amorphous carbon film of a standard TEM grid. B) Electron diffraction pattern of the region highlighted by the white circle in A. Graphite hexagonal reflections are highlighted and some crystallographic indexes are reported. The interplanar distance corresponding to the (0,-1,1,0) reflection is 0.213 nm, while the one corresponding to (-1,-1,2,0) is 0.123 nm. C) EDX spectrum acquired over the flake highlighted in A. Peaks corresponding to Si, Cu, Fe and Co comes from the TEM grid (Si and Cu) and from backscattered electrons by the polar pieces of the objective lens of the microscope.

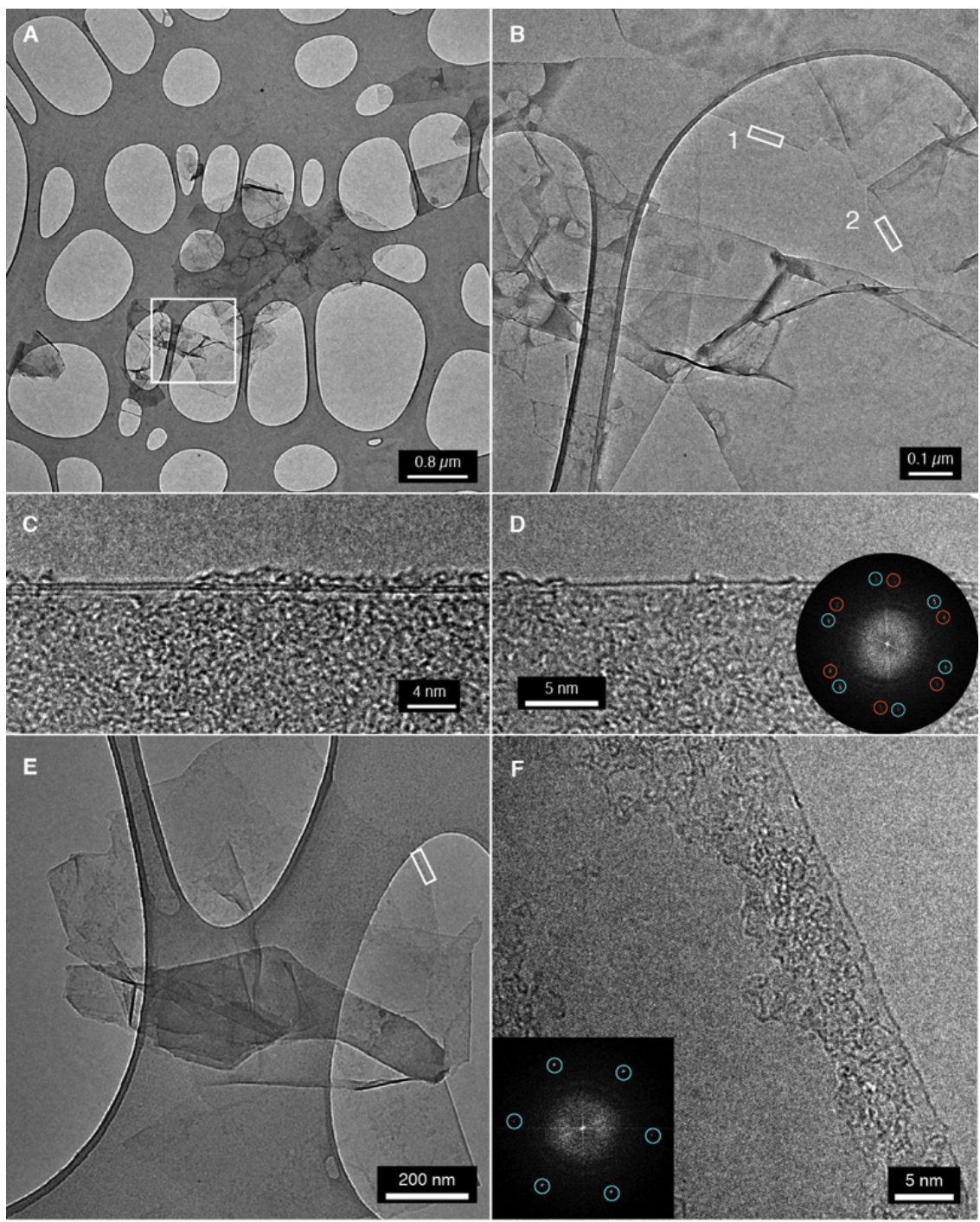

**Figure S5**: Additional results of the TEM characterization of flake thickness. A) TEM micrograph of a group of graphene flakes over the amorphous carbon film of a standard TEM grid. B) Close-up of the area highlighted by the rectangle in A. C) HRTEM image of region 1 in B, showing a bilayer fold. D) HRTEM image of the region 2 in B, showing a monolayer fold. (inset) FFT of the image, showing 2 sets of hexagonal reflections (red and blue) from the folded honeycomb lattice of the flake. E) TEM micrograph of a crumpled graphene flake over the TEM grid. F) HRTEM image of the folded monolayer edge, in the region highlighted by the white rectangle in E. (inset) FFT of the image, showing the hexagonal pattern from graphene honeycomb lattice reflections.

## 1. Graphene-graphene interaction:

Several contributions to the potential of interaction between graphene plates in water can be identified. The net balance between repulsive and attractive components will determine the stability of the graphene dispersion. Graphene flakes become electrically charged in water as a consequence of ion adsorption. The mutual repulsion due to the partial overlap of the counterion clouds is the most important contribution to the stabilizing forces. The electrostatic double layer interaction energy per unit area at distance $D$ can be estimated as [1]

$$W(D)_{Elect} = \left(64kT\rho_\infty \left(tanh(ze\psi_0/4kT)\right)^2 / \kappa \right) e^{-\kappa D}$$

where $k$ is Boltzmann's constant, $T$ the absolute temperature, $\rho_\infty$ the bulk ionic concentration, $z$ the valence of the ions in solution, $e$ the electron charge, $\psi_0$ the surface potential, and $\kappa^{-1}$ the Debye length, given by $\kappa^{-1} = (\varepsilon_0 \varepsilon kT / 2e^2 \rho_\infty z^2)^{1/2}$, where $\varepsilon$ is the dielectric constant of the medium and $\varepsilon_0$ the permittivity of free space. Calculations were performed assuming 5mM ionic concentration and $\psi_0$ equal to the measured graphene zeta potential (both conservative estimates).

The contribution of long-range undulation to the interparticle energy (which is not very significant for the case of graphene) can be estimated as [2],

$$W(D)_{Undul} \approx \left((kT)^2 / 4k_b D^2\right)$$

where $k_b$ is the bending modulus of graphene [3].

Two destabilizing contributions to the total interaction energy can be identified: van der Waals and hydrophobic interactions. The van der Waals attractive energy per unit area between infinite flat plates of thickness $a$ may be calculated as [4]

$$W(D)_{vdW} = -\frac{A_{Ham}}{12\pi}\left[\frac{1}{D^2} - \frac{2}{(D+a)^2} + \frac{1}{(D+2a)^2}\right]$$

where $A_{Ham}$ is the Hamaker coefficient for the particular combination of materials. For small separations $D$, the non-retarded, long-wavelength limit of $A_{Ham}$ can be used ($0.99 \cdot 10^{-19}$ J for graphite-graphite in water [5]). On the contrary, the distance-dependent Hamaker coefficient is necessary for larger $D$ values. We have used the $A_{Ham}$ reported by Dagastine and coworkers ([6]) for graphite in water, which is certainly a conservative choice; it has been reported that values of $A_{Ham}$ for graphene are substantially smaller than for graphite [7], in virtue of its two dimensional character. Thus, the actual repulsive barriers for graphene aggregation are likely to be larger than the ones calculated in this work. As it is not obvious what is the thickness for the transition of the optical properties of graphene to graphite, we have chosen to carry out the whole set of calculations using the Hamaker coefficients for graphite.

The hydrophobic interaction is probably the more difficult contribution to estimate; although it has often been observed that it decays exponentially with $D$, different values for the characteristic decay length can be found in the literature [8,9]. The authors of a recent experimental study of careful direct measurement of the hydrophobic interaction between fluid surfaces suggested the following approximation for the hydrophobic interaction [10],

$$W(D)_{hydrophobic} = -2\gamma e^{-D/D_0}$$

where $\gamma$ is the graphene-water interfacial energy. The most difficult parameter to estimate is the decay length of the hydrophobic interaction, $D_0$. Tabor and coworkers reported a value of 0.3 nm [10], in agreement with the Lum-Chandler-Weeks theory [11]. However, in several studies of hydrophobic interaction between solid surfaces (in thoroughly degassed conditions) $D_0$ values of the order of 1 nm have been reported [8]. We have used this value in Figures S1.

The total graphene-graphene interaction energy can then be estimated simply adding the different contributions, as

$$W(D) = W(D)_{Elect} + W(D)_{Undul} + W(D)_{vdW} + W(D)_{hydrophobic}$$

Typical results are presented in Extended Data Figure 1.

## 2. Additional Raman characterization:

**Defects** in carbon materials can be conveniently analyzed and quantified by Raman spectroscopy. Graphene in $SLG_{iw}$ can be classified as low defect « stage I » graphene according to the classification proposed by Ferrari & Robertson [12] since i) both D and G bands have narrow linewidths (27 and 21 cm$^{-1}$, respectively, at 2.33 eV) and ii) the linewidth of both D and 2D bands do not depend significantly on laser energy (Table S1) [13,14,12]. Furthermore, a comparison of the double resonant defect-induced bands D and D' can provide information on the nature of defects [15]. It has been shown, at 2.4 eV excitation energy, that $I_D/I_{D'}$ = 13 for *sp³* type defects, $I_D/I_{D'}$ = 7 for vacancy defects, whilst $I_D/I_{D'}$ = 3.5 for edge defects (measured on polycrystalline graphite) [15]. SLG$_{iw}$ shows a ratio $I_D/I_{D'}$ = 9 at 2.33 eV. We attribute this result to the coexistence of edge defects and *sp³* defects, likely due to some functionalization of the flakes with -OH or -H groups [16][17]). The typical distance between defects, $L_d$, can be estimated from the linewidth of the main bands, and the $I_D/I_G$ ratio [18,19,14]. Following the analysis of ref [19] and considering that the structural radius for the sp³ defects lies between the carbon-carbon distance (0.142 nm) and 1 nm [13,20], we find a typical distance between defects in the range 7-10 nm, which corresponds to a concentration of defects in the range of 300-600 ppm (detailed calculation in the next paragraph)

| Excitation Energy (eV) | D | | G | | D' | | 2D | | $I_D/I_G$ | $I_D/I_{D'}$ | $I_{2D}/I_G$ |
|---|---|---|---|---|---|---|---|---|---|---|---|
| | ω | 2Γ | ω | 2Γ | ω | 2Γ | ω | 2Γ | | | |
| 2.33 | 1345 | 27 | 1586 | 21 | 1620 | 16 | 2681 | 28 | 1.5 | 9.0 | 2.0 |
| 1.94 | 1325 | 26 | 1585 | 23 | 1617 | 17 | 2643 | 30 | 2.7 | 6.9 | 2.7 |
| 1.58 | 1303 | 28 | 1586 | 23 | 1612 | 18 | 2599 | 27 | 4.4 | 6.6 | 2.3 |
| 1.17 | 1277 | 34 | 1586 | 25 | 1605 | 19 | 2543 | 28 | 4.2 | 5.5 | * |

**Table S1**. Position ω (cm⁻¹), linewidth Γ (cm⁻¹) and relevant intensity ratios as a function of excitation energy. * $I_{2D}/I_G$ could not be measured properly at 1.17 eV because of strong absorption bands of water in the near infrared.

**Estimation of the defect density**

In the activation radius model, initially proposed by Lucchese *et al* [21] and developed by Cançado *et al* [19], a point defect is associated to a structural radius $r_S$, corresponding to the area where the structure is changed, and an activated radius $r_A$, corresponding to the area where the Raman D band is activated. In this model, in the limit of low defect density (with a typical distance between defects $L_D > 10$ nm),

$$L_D^2 \approx C_A . \pi (r_A^2 - r_S^2) \left(\frac{I_D}{I_G}\right)^{-1}$$

where $C_A$ corresponds to the maximum of $I_D/I_G$, and is therefore independent of the nature of defects, and is expressed as $C_A \approx 160 E_L^{-4}$ where $E_L$ is the exciting laser energy. The relationship between $L_D$ and $I_D/I_G$ includes correcting terms for $L_D<10$ nm (equation 1 in reference [19]), which can be neglected in first approximation for $L_D>7$ nm. For Ar⁺ bombardment-induced vacancy defects, $r_A=3$ nm and $r_S=1$ nm [21,19]. For sp³ point defects, $r_S$ could be smaller, but not smaller than the C-C distance, *i.e.* ≈ 0.142 nm. Note that several groups considered that $r_S$ should be close for vacancies and sp$^3$ defects [13],[20]. On the other hand, ($r_A$-$r_S$) corresponds to the correlation length of photoexcited electrons participating in the double-resonance mechanism responsible for the D band, and should be very close for all point defects. Therefore, the relation above, or in an equivalent way equation 1 in ref [19], can be used to estimate $L_D$ from Raman measurements on SLG$_{iw}$. From our data measured with three different laser lines, we find $L_D$=7.5±1 nm if we take $r_S$=0.142 nm, and $L_D$=10±1 nm if we take $r_S$=1 nm. This is in good agreement with $L_D$ estimated from the linewidth of the main bands. This leads to a defect density in the range 300-600 ppm, very close to that estimated by Hirsch et al in reference [20].

## 3. Detailed conductivity data.

Conductivity of the films were measured by the 4 point method after evaporating gold contacts onto the films (Figure S6). A typical I-V curve is represented on Figure S7. Results are summarized in Table S2.

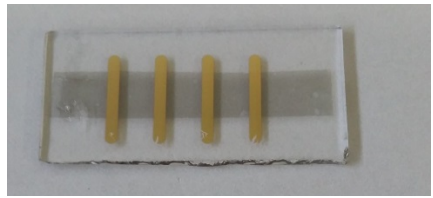

Figure S6. Film with gold contacts evaporated on it. The distance between contacts is 4 mm.

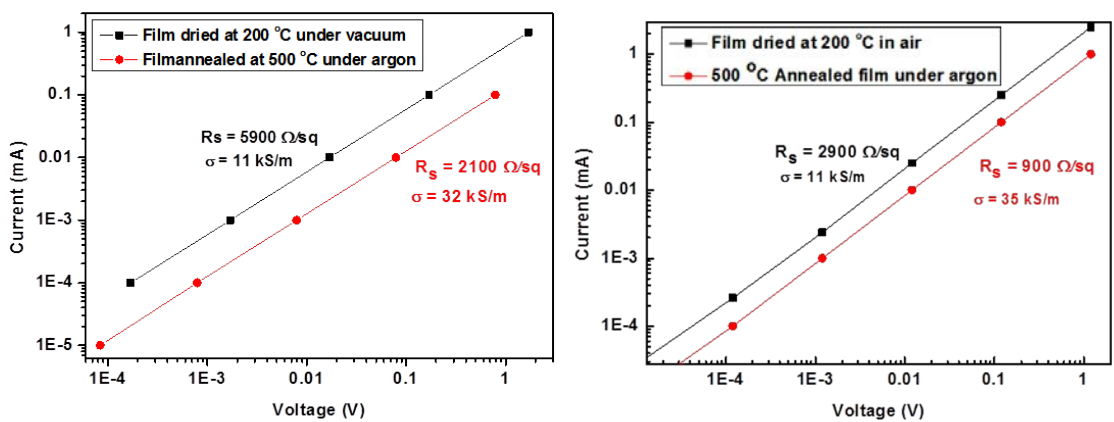

Figure S7. Typical I-V curves obtained. Transmittance of these specific films were 65 % (left) and 35% (right).

| Device | Thickness (nm) | Rs (Ω/□) 200 °C in Vac | σ (S/m) 200 in Vac | Rs (Ω/□) 500 °C in Ar | σ (S/m) 500 in Vac |
|---|---|---|---|---|---|
| 1 | 15 | 5900 | 11000 | 2100 | 32000 |
| 2 | 15 | 9800 | 6800 | 3500 | 19000 |
| 3 | 15 | 17000 | 4000 | 5500 | 12000 |
| 4 | 15 | 15000 | 4400 | 4700 | 14000 |
| 5 | 35 | 3100 | 9200 | 820 | 35000 |
| 6 | 30 | 2900 | 11000 | 900 | 35000 |

Table S2. Thickness, surface resistance and conductivity for different films prepared from $SLG_{iw}$ Average resulting conductivities are 7000 S/m and 20000 S/m after drying respectively at 200°C in vacuum and 500 °C under argon. The 15 nm films have a transmittance of 65 %.

## 4. X-ray Photoelectron Spectroscopy Analysis.

C1s spectrum of a graphene film prepared from $SLG_{iw}$ is shown in Figure S8 after drying in vacuum at room temperature and after annealing at 500 °C with comparison with graphite and a graphene film prepared by directly filtering the graphenide solution in THF. One can see that the higher energy side, indicative of sp3 carbon functionalization remains small in all cases and that annealing lowers it to the level of the film obtained from graphenide solutions

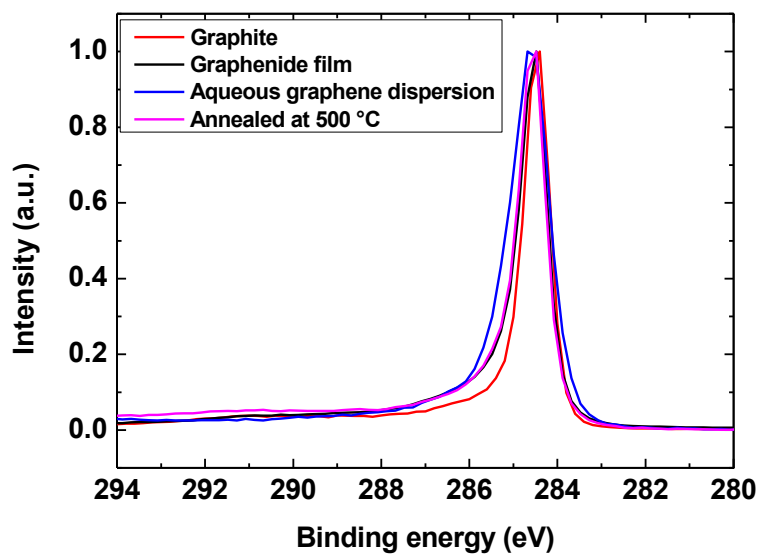

Figure S8. C1S spectrum of graphite (red), a graphene film obtained by filtering a graphenide solution in THF, then exposing it to air (black), a graphene film obtained from $SLG_{iw}$ (blue) and the same film after annealing at 500 °C. RGO films, by contrast, show a series of higher energy peaks between 285 and 292 eV eV attributed to functionalized sp3 carbon atoms[22].

**5. Comparative opacity of SLG$_{iw}$ dispersion and sodium cholate FLG dispersions.**

SLG$_{iw}$ dispersions carefully adjusted at the same concentration as a sodium cholate FLG dispersion, show much higher transparency (Figure S9), indicating a significant difference in the nature of the two dispersions. We are now investigating the possible physical reasons for this phenomenon.

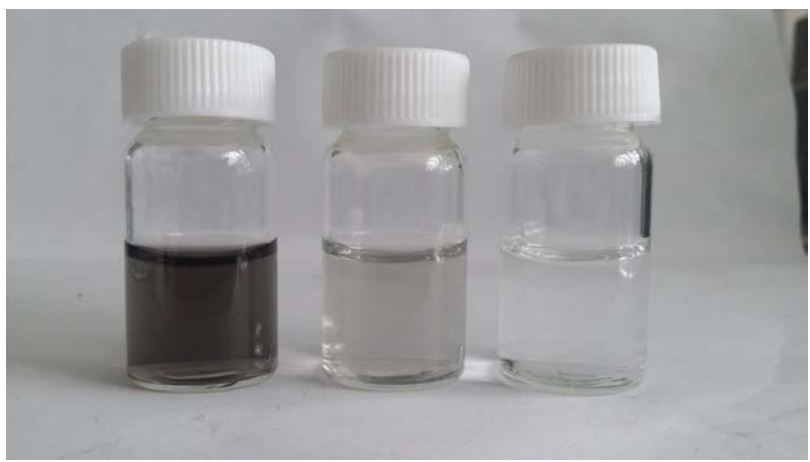

Figure S9. Vials of Na cholate few layer graphene dispersion in water at 0.08 mg graphene/ml (left), SLG$_{iw}$ dispersion at 0.08 mg graphene/ml (middle) and pure water (right).

Experimental procedure: The concentration of a FLG dispersion was determined by filtering a known volume of sodium cholate stabilized graphene dispersion through a nitrocellulose filter membrane (milipore, 25 nm pore size) and washing with copious amounts of water to remove sodium cholate. The nitrocellulose membrane now containing graphene was re-weighed after drying in a vacuum oven at 60 ° C for 12 hours. The sodium cholate stabilized FLG dispersion contained 0.12 mg/mL of graphene. Concentration of an SLG$_{iw}$ sample (0.08 mg/mL) was determined using the same vacuum filtration procedure (during which the residual KOH was eliminated by the rinsing step. No residual potassium could be detected by XPS analysis). The concentration of the sodium cholate stabilized graphene dispersion was confirmed by freeze drying a known volume and weighing the remaining material. Residual mass of sodium cholate in the dried material was taken into consideration.

The sodium cholate stabilized FLG dispersion (0.12 mg/mL) was diluted to match the concentration of the SLG$_{iw}$ sample (0.08 mg/mL).

**References and Notes:**

1. Israelachvili, J. N. in *Intermol. Surf. Forces* 205–222 (Elsevier, 2011). doi:10.1016/B978-0-12-375182-9.10011-9

2. Servuss, R. M. & Helfrich, W. Mutual adhesion of lecithin membranes at ultralow tensions. *J. Phys.* **50,** 809–827 (1989).

3. Lu, Q., Arroyo, M. & Huang, R. Elastic bending modulus of monolayer graphene. *J. Phys. D. Appl. Phys.* **42,** 102002 (2009).

4. Parsegian, V. A. *Van der Waals Forces A Handbook for Biologists, Chemists, Engineers, and Physicists*. (Cambridge University Press, 2005).

5. Li, J.-L. *et al.* Use of dielectric functions in the theory of dispersion forces. *Phys. Rev. B* **71,** 235412 (2005).

6. Dagastine, R. R., Prieve, D. C. & White, L. R. Calculations of van der Waals forces in 2-dimensionally anisotropic materials and its application to carbon black. *J. Colloid Interface Sci.* **249,** 78–83 (2002).

7. Rajter, R. F., French, R. H., Ching, W. Y., Carter, W. C. & Chiang, Y. M. Calculating van der Waals-London dispersion spectra and Hamaker coefficients of carbon nanotubes in water from ab initio optical properties. *J. Appl. Phys.* **101,** 17–20 (2007).

8. Meyer, E. E., Rosenberg, K. J. & Israelachvili, J. Recent progress in understanding hydrophobic interactions. *Proc. Natl. Acad. Sci. U. S. A.* **103,** 15739–15746 (2006).

9. Hammer, M. U., Anderson, T. H., Chaimovich, A., Shell, M. S. & Israelachvili, J. The search for the hydrophobic force law. *Faraday Discuss.* **146,** 299 (2010).

10. Tabor, R. F., Wu, C., Grieser, F., Dagastine, R. R. & Chan, D. Y. C. Measurement of the Hydrophobic Force in a Soft Matter System. *J. Phys. Chem. Lett.* **4,** 3872–3877 (2013).

11. Chandler, D. Interfaces and the driving force of hydrophobic assembly. *Nature* **437,** 640–7 (2005).

12. Ferrari, a. & Robertson, J. Interpretation of Raman spectra of disordered and amorphous carbon. *Phys. Rev. B* **61,** 14095–14107 (2000).

13. Eckmann, A., Felten, A., Verzhbitskiy, I., Davey, R. & Casiraghi, C. Raman study on defective graphene: Effect of the excitation energy, type, and amount of defects. *Phys. Rev. B* **88,** 035426 (2013).

14. Martins Ferreira, E. H. *et al.* Evolution of the Raman spectra from single-, few-, and many-layer graphene with increasing disorder. *Phys. Rev. B* **82,** 125429 (2010).

15. Eckmann, A. *et al.* Probing the Nature of Defects in Graphene by Raman Spectroscopy. *Nano Lett.* (2012). doi:10.1021/nl300901a

16. Hof, F., Bosch, S., Eigler, S., Hauke, F. & Hirsch, A. New Basic Insight into Reductive Functionalization Sequences of Single Walled Carbon Nanotubes (SWCNTs). *J. Am. Chem. Soc.* **135,** 18385–18395 (2013).

17. Schäfer, R. a *et al.* On the way to graphane-pronounced fluorescence of polyhydrogenated graphene. *Angew. Chem. Int. Ed. Engl.* **52,** 754–7 (2013).

18. Ferrari, A. C. & Basko, D. M. Raman spectroscopy as a versatile tool for studying the properties of graphene. *Nat. Nanotechnol.* **8,** 235–46 (2013).

19. Cançado, L. G. *et al.* Quantifying defects in graphene via Raman spectroscopy at different excitation energies. *Nano Lett.* **11,** 3190–6 (2011).

20. Eigler, S. & Hirsch, A. Chemistry with Graphene and Graphene Oxide-Challenges for Synthetic Chemists. *Angew. Chemie Int. Ed.* **53,** 7720–7738 (2014).

21. Lucchese, M. M. *et al.* Quantifying ion-induced defects and Raman relaxation length in graphene. *Carbon N. Y.* **48,** 1592–1597 (2010).

22. Becerril, H. A. *et al.* Evaluation of Solution-Processed Reduced Graphene Oxide Films as Transparent conductors. *ACS Nano.* **2,** 463–470 (2008).